\documentclass[cits]{PoS}
\usepackage[cspex,bbgreekl]{mathbbol}
\title{Hunting for the strangeness content of the nucleon}

\ShortTitle{Hunting for the strangeness content of the nucleon}

\author{\speaker{Gunnar Bali}, Sara Collins and Andreas Sch\"afer\\
        Institut f\"ur Theoretische Physik, Universit\"at Regensburg,\\
        93040 Regensburg, Germany\\
        E-mail:\\\email{gunnar.bali@physik.uni-regensburg.de},
\email{sara.collins@physik.uni-regensburg.de},
\email{andreas.schaefer@physik.uni-regensburg.de}}

      \abstract{We present results for the strangeness contribution to
        the nucleon, $\langle N|\bar{s}s|N\rangle$ and to the spin of
        the nucleon, $\Delta s$. By combining several variance
        reduction techniques for all-to-all propagators we are able
        to obtain gains in terms of computer time of factors of
        25--30 for the disconnected loop that is needed within
        the calculation of $\Delta s$, relative to the standard approach
        of just employing
        time partitioning/dilution . For
        $\langle N|\bar{s}s|N\rangle$, the error is dominated by the
        gauge noise.}

\FullConference{The XXVI International Symposium on Lattice Field Theory\\
                 July 14 - 19, 2008\\
                 Williamsburg, Virginia, USA}
\newcommand{\dslash}{\! \not \!\! D}
\begin{document}

\section{Introduction}
Many nucleon structure observables require the calculation of
disconnected quark line diagrams for which all-to-all
propagator techniques are needed. Here we present first results of an
ongoing project to
calculate the strangeness contribution to the spin
of the nucleon $\Delta s$ as well as the scalar strangeness
content of the nucleon $\langle N|\bar{s}s|N\rangle$, using improved
stochastic methods.

The spin of the nucleon can be factorized into
a quark spin contribution
$\Delta\Sigma$, a quark angular momentum contribution $L_q$
and a gluonic contribution (spin and angular momentum) $\Delta G$:
\begin{equation}
\frac12=\frac12 \Delta\Sigma+L_q+\Delta G\,.
\end{equation}
In the na\"{\i}ve $SU(6)$ quark model, $\Delta \Sigma=1$, with
vanishing angular momentum and gluon contributions. In this
case sea quark contributions will be absent too and
therefore there will be no strangeness
contribution $\Delta s$ in the factorisation,
\begin{equation}
\Delta\Sigma=\Delta d+\Delta u +\Delta s+\cdots\,,
\end{equation}
where in our notation $\Delta q$ contains both, the spin of
the quarks $q$ and of the antiquarks $\bar{q}$.
Experimentally $\Delta s$ is usually
obtained by integrating the strangeness contribution to
the spin structure function $g_1$ over momentum transfers $x$.
The integral over the range in which data exists ($x\gtrsim 0.004$)
typically agrees with zero which means that a non-zero result
relies on the unprobed very small-$x$ region and is
model dependent. Recent Hermes analysis~\cite{Airapetian:2007mh}
yields $\Delta s=-0.085(13)(8)(9)$
at a renormalization
scale $\mu^2=5\,$GeV${}^2$ in the $\overline{MS}$ scheme
while our (as yet unrenormalized) results suggest
$|\Delta s|<0.01$.

The scalar strangeness density is not directly accessible in
experiment but plays a r\^ole in models of nuclear structure.
It is also of phenomenological interest since,
assuming that heavy flavours are strongly suppressed,
the dominant coupling of the Higgs particle to the nucleon will
be accompanied by this scalar matrix element.

We will first discuss our methods, then
the error reduction achieved in
our present lattice setup and finally we present results on
the two matrix elements, before concluding.

\section{Stochastic methods}
We denote the lattice spacing by $a$
and the lattice Dirac matrix by $M={\mathbb 1}-\kappa\dslash$.
Disconnected quark line contributions
require all-to-all propagators $M^{-1}_{ji}$ where the
multi-index $i=(x,\alpha,a)$ runs over all colours
$a=1,2,3$, spinor indices $\alpha=1,\ldots,4$ and
spacetime sites $x\in V$. Note that in our particular application
it is natural and sufficient to restrict $x$ to a given timeslice.
Exact methods to obtain $M^{-1}$ are unfeasible in terms
of computer time and memory since $12 V$ solver applications are required.
Employing stochastic methods~\cite{Bitar:1988bb}, this factor can be substituted
by the number of estimates $L\ll 12 V$: in a first step
a set of Dirac noise vectors 
$\{|\eta_{\ell}\rangle:\ell=1, \ldots, L\}$ is generated
where the $12 V$ complex colour-spinor-site components are
filled with $({\mathbb Z}_2\otimes i\,{\mathbb Z}_2)/\sqrt{2}$
uncorrelated random numbers~\cite{Dong:1993pk}.
These have the following properties:
\begin{equation}
\overline{ |\eta\rangle\langle\eta|}_L:=
\frac{1}{L}\sum_{\ell}|\eta_{\ell}
\rangle \langle \eta_{\ell}|
= \mathbb{1} + \mathcal{O}(1/\sqrt{L})\,,\qquad
\overline{ \langle \eta|} = \mathcal{O}(1/\sqrt{L})\,.
\end{equation}
We will also employ the short-hand notation
$\overline{|\cdot\rangle\langle\cdot|}
=\overline{|\cdot\rangle\langle\cdot|}_L$.
We use the conjugate gradient algorithm with even/odd preconditioning
to obtain the
solutions $|s_{\ell}\rangle$ of the sparse linear problems,
\begin{eqnarray}
M|s_{\ell}\rangle & = & |\eta_{\ell} \rangle\,.
\end{eqnarray}
From these one can construct an unbiased estimate of $M^{-1}$:
\begin{equation}
\label{eq:esti}
E(M^{-1}):= \overline{|s\rangle \langle \eta|}
             = M^{-1} +  M^{-1}\underbrace{(\overline{| \eta\rangle\langle \eta|}-
\mathbb{1})}_{\mathcal{O}(1/\sqrt{L})}\,.
\end{equation}
Due to the difference between $E(M^{-1})$ and $M^{-1}$ above,
any fermionic observable $A$ can only be estimated up to
a stochastic error $\Delta_{\rm stoch}A=\mathcal{O}(1/\sqrt{L})$
on a given configuration. We define the configuration
average $\langle\cdot\rangle_c$
over $n_{\rm conf}$ uncorrelated configurations 
and normalize this
appropriately:
\begin{equation}
\label{eq:sto}
\sigma_{A,\rm stoch}^2:=
\frac{\langle\Delta_{A,\rm stoch}^2\rangle_c}{n_{\rm conf}}\,.
\end{equation}
For large $L$ and $n_{\rm conf}$ this will scale like
$\sigma_{A,\rm stoch}^2\propto (Ln_{\rm conf})^{-1}$.
We also define the gauge error
$\sigma^2_{A,\rm gauge}\propto n_{\rm conf}^{-1}$
as the variation of the estimates of $A$ over gauge
configurations. This will be minimized at fixed
$n_{\rm conf}$ if $A$ is calculated
exactly. In general the gauge error is limited by, 
\begin{equation}
\sigma^2_{A,\rm gauge}\geq\sigma^2_{A,\rm stoch}\,.
\end{equation}
If $\sigma^2_{A,\rm stoch}\simeq\sigma^2_{A,\rm gauge}$
then obviously it is worthwhile to improve the quality
of the estimates while if $\sigma^2_{A,\rm stoch}\ll
\sigma^2_{A,\rm gauge}$ then precision can only
be gained by increasing $n_{\rm conf}$, possibly
reducing $L$ to save computer time since the
same $n_{\rm conf}^{-1}$ scaling enters both
sides of the inequality.

In our calculation of $\Delta s$ the stochastic error initially
was dominant. Hence we
combined several variance reduction techniques to
reduce this:
\begin{itemize}
\item partitioning (also coined dilution)~\cite{Bernardson:1993yg}:
we only set $|\eta_{\ell}\rangle\neq 0$
on one timeslice. This removes some of the (larger) off-diagonal
noise elements, see eq.~(\ref{eq:esti}), and reduces the variance.
\item hopping parameter expansion (HPE)~\cite{hopex}: the first
few terms
of the hopping parameter expansion
of $\mbox{Tr}(\Gamma M^{-1})=\mbox{Tr}[\Gamma
(\mathbb{1}-\kappa\dslash)^{-1}]$ vanish identically but still
contribute to the noise. For the Wilson action,
$\mathrm{Tr}(\Gamma M^{-1})=
\mathrm{Tr}(\Gamma\kappa^{n}\dslash^{n}{M}^{-1})$ for
$n=4, 8$, depending on $\Gamma$, where for $\Gamma={\mathbb 1}$
one can easily calculate and correct for the zero-order difference.
\item truncated solver method (TSM)~\cite{lat07}:
calculate approximate solutions $|s_{n_{\rm t},\ell}\rangle$ after
$n_{\rm t}$ solver iterations (before
convergence), and
estimate the difference stochastically to obtain an
unbiased estimate of $M^{-1}$:
\[
 \mathrm{E}(M^{-1})=
\overline{|s_{n_{\rm t}}\rangle\langle\eta|}_{L_1}+
\overline{(|s\rangle-|s_{n_{\rm t}}\rangle)\langle\eta|}_{L_2}\quad
\mbox{where}\quad L_2\ll L_1\,.
\]
\item Truncated eigenmode approach (TEA)~\cite{eigen,stringbreak}: calculate the
$n_{\rm ev}$ lowest eigenvalues and eigenvectors of
$\mathrm{Q}=\gamma_5M=Q^{\dagger}$, $
\mathrm{Q}^{-1}  =  \mathrm{Q}^{-1}_{\perp}
 +  \sum_{i=1}^{n_{\rm ev}}|u_i\rangle q_i^{-1}\langle u_i|$,
and stochastically estimate the complement
$\mathrm{Q}^{-1}_{\perp}$ (with deflation included for free).
\end{itemize}

\section{Lattice setup and error reduction}
Our exploratory calculations are performed on $V=16^3\times 32$
configurations of $n_f\approx 2+1$ rooted stout-link
improved staggered quarks with a Symanzik
improved gauge action. These were
provided by the Wuppertal group.
The lattice spacing is fairly
coarse, $a^{-1}\approx 1.55$~GeV, and the spatial dimension
is around $2$~fm~\cite{latdetails}.
We used the Wilson action for our valence quarks and
currents with $\kappa=0.166$, $0.1675$ and $0.1684$,
corresponding to pseudoscalar masses of about $600$, $450$ and
$300$~MeV respectively. The analysis was performed on 326
configurations at $\kappa_{\rm loop}=0.166$, 167 configurations at
$\kappa_{\rm loop}=0.1675$ and $152$ configurations at
$\kappa_{\rm loop}=0.1684$, where $\kappa_{\rm loop}$
refers to the $\kappa$
value of the disconnected loop. Throughout we used a
modified version of the Chroma code~\cite{chroma}.

On each configuration the disconnected loop was calculated using
the stochastic variance reduction techniques detailed
above (the TEA was only used at
$\kappa_{\rm loop}=0.1684$, where $20$ eigenvalues
were calculated).
We investigate the reduction in computer time, using
optimized stochastic estimates, relative to those
without any improvement techniques
applied (except for time partitioning).
We state
all costs in terms of the average {\em real} computer time required
on a Pentium 4 PC for one solver application (unimproved estimate),
where we account for all overheads of the improvement methods.

\TABULAR{|l|l|r|l|l|l|l|}
{\hline
$\mathrm{Tr}(\Gamma_{\rm loop}\mathrm{M}^{-1})$ &$\kappa_{\rm loop}$ &cost & loop$^{\rm opt}$ & $\sigma_{\rm stoch}^{\rm opt}$ &loop & $\sigma_{\rm stoch}$ \\\hline
$\Gamma_{\rm loop} =\frac{1}{3}\sum_j\gamma_j\gamma_5$ &  0.166 &
 300  & -0.008(50)  & 0.016  &  &  \\
& & 100 & -0.033(55)& 0.027  & -0.185(148) & 0.135  \\
& & 50 & -0.054(64)& 0.039 & -0.446(201)  & 0.186 \\\hline
 & 0.1675 &300 & -0.085~~(87) & 0.030 & & \\
& & 100 & -0.040(101) & 0.054 &0.003(211) & 0.198 \\
& &  50 & -0.038(114) & 0.076 &0.056(265) & 0.271 \\\hline
& 0.1684 & 300 & -0.069(95) & 0.015  & &\\
&        & 100 & -0.068(96) &0.036  &-0.089(216) & 0.212\\\hline\hline
$\Gamma_{\rm loop}=\mathbb{1}$ & 0.166  & 300  & 14702.6(7) & 0.04   & &\\
                    &        &  12  & 14702.5(7) & 0.18  & 14703.5~~~(9) & 0.47\\
                    &        &   6  & 14702.3(8) & 0.23  & 14703.7(1.0)& 0.65\\\hline
                    &  0.1675 & 300 & 14743.1(1.1) & 0.06 & &\\
                    &         &  12 & 14743.4(1.2) & 0.33 & 14745.0(1.3) & 0.69\\
                    &         &   6 & 14743.5(1.2) & 0.42 & 14744.6(1.5) & 0.96\\\hline
&0.1684 &300 & 14764.9(1.2) & 0.04 & &\\
&       &100 & 14764.9(1.2) & 0.08 &14764.6(1.2) & 0.27\\\hline}
{Results for the disconnected loop, averaged over configurations,
obtained with~(loop$^{\rm opt}$) and without~(loop) variance reduction
techniques. The cost is in units of the average computer time required
to solve for one (undeflated) right hand side.\label{tab1}}

Results for the configuration averages of the loops
$\mathrm{Tr}(\Gamma_{\rm loop}\mathrm{M}^{-1})$
are given in table~\ref{tab1}. The gauge errors $\sigma_{\rm gauge}$
(that also depend on the stochastic noise) are displayed
in brackets after the loop averages.
These can be compared to the
purely stochastic errors $\sigma_{\rm stoch}$,
defined in eq.~(\ref{eq:sto}).
The deflation at $\kappa_{\rm loop}=0.1684$ where we
apply TEA accelerates the solver but time is required
for the eigenvector set-up. In our implementation
the cost of solving for about 90 undeflated right hand sides
equals that of 90 deflated ones (including this overhead).
This is why in this case we do not display results obtained at
the lower cost values.

For
$\mathrm{Tr}(\frac{1}{3}\sum_j \gamma_j \gamma_5\mathrm{M}^{-1})$ the
stochastic error dominates over the gauge error unless $L$ is chosen
ridiculously large or variance reduction
techniques are applied. Using these
techniques the error is brought under control to the extent that 
we only need to invest the computer time equivalent of
roughly 100 unimproved stochastic estimates to achieve
$\sigma_{\rm stoch}<\frac12\sigma_{\rm gauge}$.
In particular, we find a reduction in $\sigma_{\rm stoch}^2$~(which is
proportional to the amount of computer time required) of approximately
25--30 for $\kappa_{\rm loop}=0.166$ and $0.1684$. A
smaller gain is
obtained for the intermediate
$\kappa_{\rm loop}=0.1675$ which may benefit from using the TEA approach.
For $\mathrm{Tr}(\mathbb{1}\mathrm{M}^{-1})$ the situation is
reversed and the gauge error clearly
dominates over the stochastic error: apart from possibly the
heaviest $\kappa_{\rm loop}$ there is no advantage in using variance
reduction techniques. 

The matrix elements,
\begin{equation}
\langle N,s|\bar{q}\gamma_{\mu}\gamma_5 q|N,s\rangle = 2M_N s_{\mu}
\frac{\Delta q}{2}
\end{equation}
and $\langle N|\bar{q}q|N\rangle$ are extracted from the ratios of
three-point functions to two-point functions~(at zero momentum):
\begin{equation}
R^{\rm dis}(t,t_f) = 
-\frac{\langle\Gamma_{\rm 2pt}^{\alpha\beta}C^{\beta\alpha}_{\rm 2pt}(t_0,t_f) \sum_{\mathbf{x}}\mathrm{Tr}(\Gamma_{\rm loop} M^{-1}(\mathbf{x},t;\mathbf{x},t))\rangle}{\langle \Gamma_{\rm unpol}^{\alpha\beta} C^{\beta\alpha}_{\rm 2pt}(t_0,t_f)\rangle}
\end{equation}
where $\Gamma_{\rm 2pt}= \Gamma_{\rm unpol} =(1+\gamma_4)/2$ and
$\Gamma_{\rm loop}=\mathbb{1}$ for $\langle N|\bar{q}q|N\rangle$ and
$\Gamma_{\rm 2pt}= i\gamma_j\gamma_5(1+\gamma_4)/2$ and
$\Gamma_{\rm loop}=\gamma_j\gamma_5$ for $\Delta q$, where we
average over $j=1,2,3$. Note that for
$q=u,d$ there is an additional connected contribution $R^{\rm con}$, which
we have not calculated. We combine the three $\kappa_{\rm loop}$ values
with $\kappa_{\rm 2pt}=0.166$ and 0.1675.
In the limit of large times, $t_f\gg t\gg t_0$,
\begin{equation}
R^{\rm dis}(t,t_f)+R^{\rm con}(t,t_f) \rightarrow
2\frac{\langle N,s|(\bar{q}\Gamma_{\rm loop}q)^{\rm latt}|N,s\rangle}{2M_N}\,.
\end{equation}

\FIGURE
{
\rotatebox{270}{\includegraphics[height=.45\textwidth,clip]{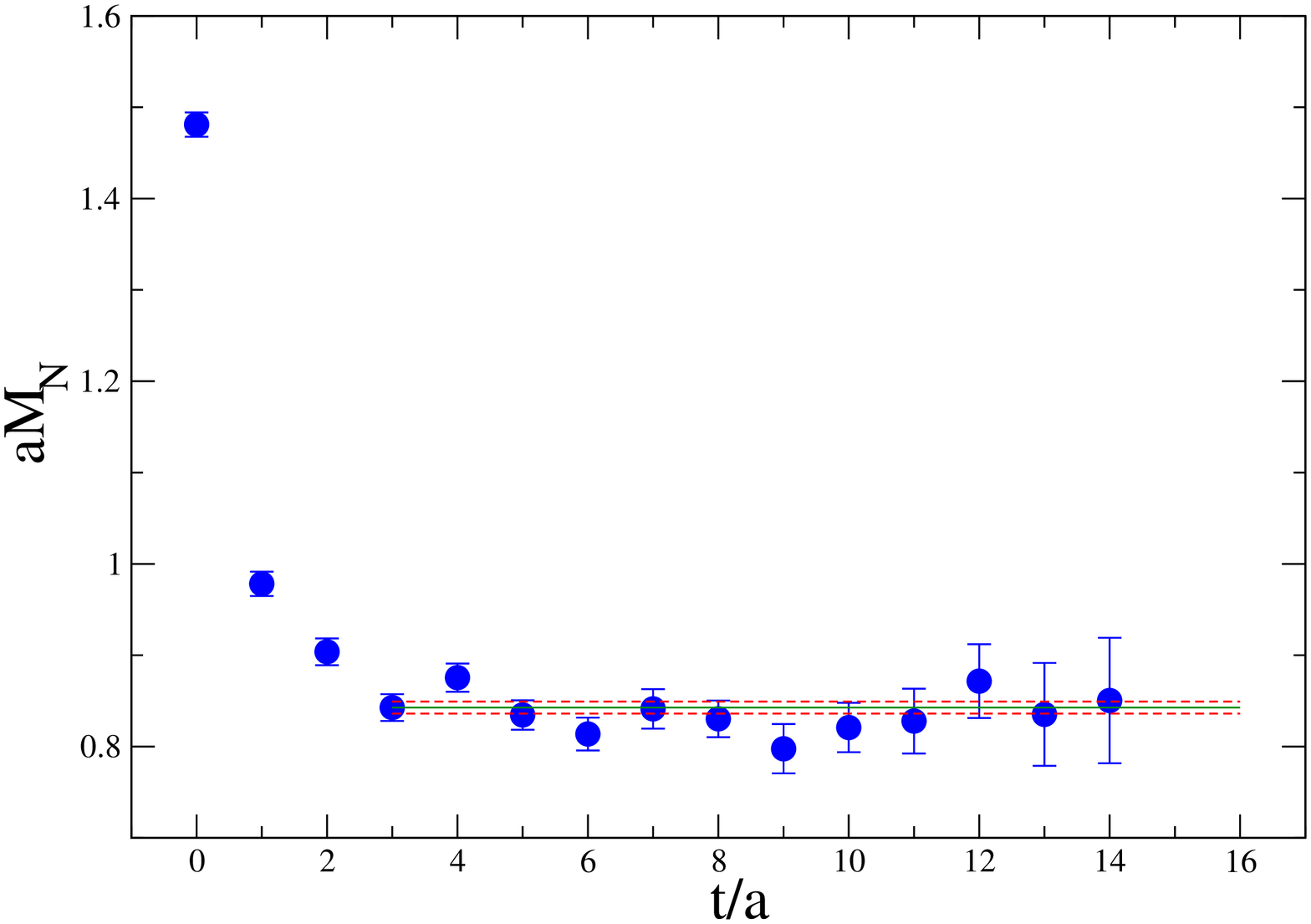}}\hspace*{.05\textwidth}
\rotatebox{270}{\includegraphics[height=.45\textwidth,clip]{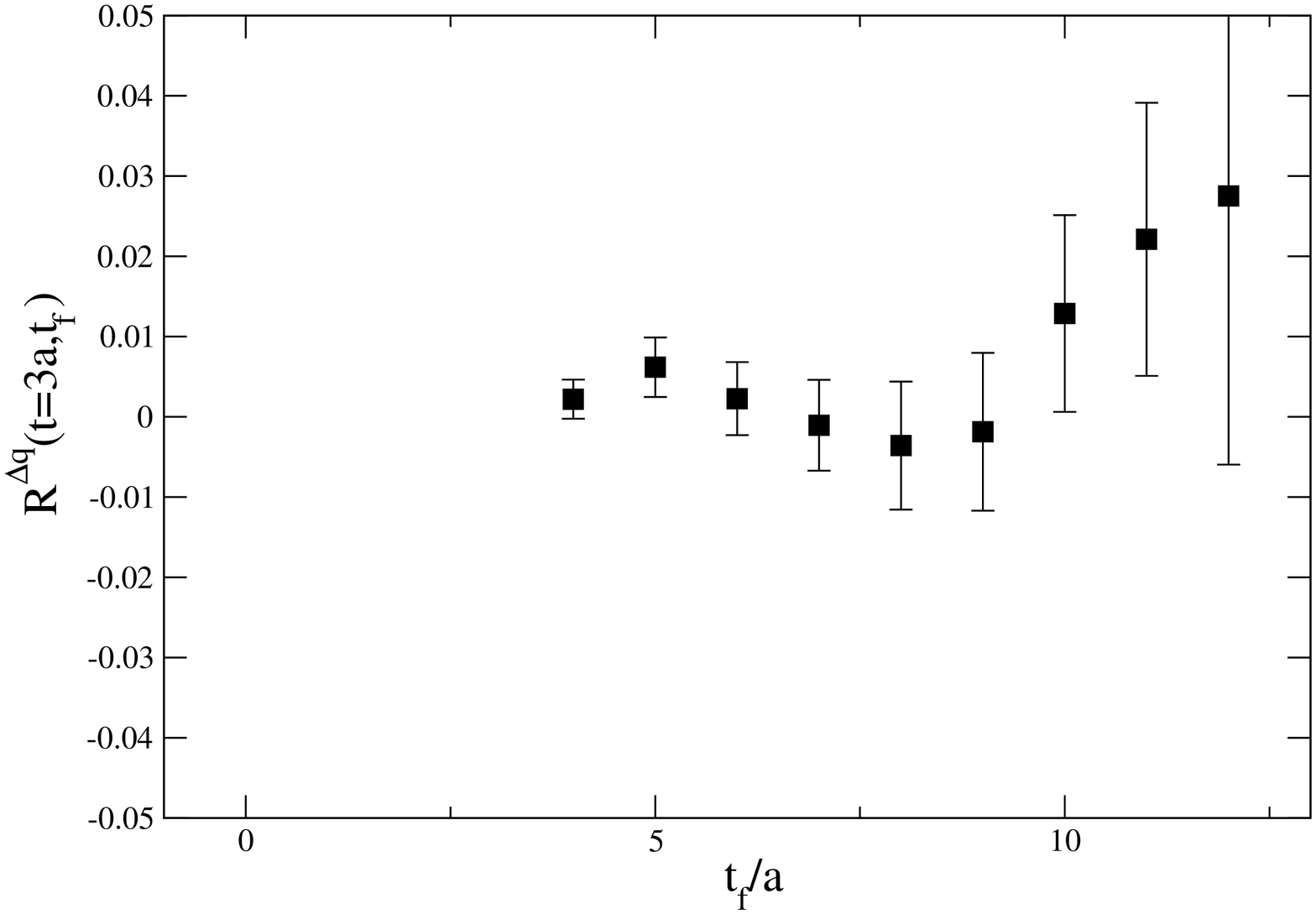}}
\caption{
The effective mass of the proton with
$\kappa_{\rm 2pt}=0.166$ (left).
The ratio, $R^{\Delta q}(t=3a,t_f)$ as a function
  of $t_f$ for $\kappa_{\rm loop}=\kappa_{\rm 2pt}=0.166$ (right).
\label{fig2}}}

We optimized the
nucleon creation and
annihilation operators
using Wuppertal smearing with spatial APE-smeared parallel
transporters~\cite{stringbreak}.
The effective mass plot of figure~\ref{fig2} illustrates
ground state dominance from a time $t=3a\approx 0.38\,$fm onwards.
The same holds for $\kappa_{\rm 2pt}=0.1675.$
Hence we place the source at $t_0=0$, the current insertion
at $t=3a$ and destroy the nucleon at $t_f\ge 4a$.
The result on the right of figure~\ref{fig2}
does not depend on $t_f$, even for $t_f<6a$, indicating that indeed
with the chosen temporal separations we effectively realize the large-$t$
limit. In table~\ref{tab2} we display the
results for $\Delta q^{\rm dis}$ at the symmetric point
$t_f=6a\approx 0.76$~fm:
our methods enable us to reduce the squared errors by
factors ranging from 5.5 to 11 at the fixed computational cost of
100 solver applications (in addition to calculating the two-point function).
This falls somewhat short of the gains
that we achieved in table~\ref{tab1} for the loops alone since
now there are additional sources of gauge error. These we attempt to address
in the near future.

\TABULAR{|r|c|c||c|c||c|c|}
{\hline 
&\multicolumn{2}{c||}{$\kappa_{\rm loop}=0.166$} &
 \multicolumn{2}{c||}{$\kappa_{\rm loop}=0.1675$}&
 \multicolumn{2}{c|} {$\kappa_{\rm loop}=0.1684$}\\\hline
 \multicolumn{7}{|c|}{$\kappa_{\rm 2pt}=0.166$}\\\hline
cost     &  R$^{\rm opt}$ & R  & R$^{\rm opt}$ & R&R$^{\rm opt}$&R \\\hline
300&-0.001(4)&             & -0.002~~(7) &            &-0.001~~(7)   &             \\
100&-0.002(5)& +0.005(14)& -0.001~~(9)&+0.008(22)&-0.004~~(7)   & +0.008(20)\\
 50&+0.001(6)& +0.021(17)& +0.004(10)&+0.036(27)&&\\\hline
  \multicolumn{7}{|c|}{$\kappa_{\rm 2pt}=0.1675$}\\\hline
300 &-0.005(6)   &             &-0.003(12)&            &-0.004(13)&            \\
100 &-0.008(7)   & +0.009(23)&+0.005(15)&+0.028(35)&-0.006(13)&-0.004(28)\\
 50 &-0.002(9)   & +0.046(29)&+0.023(17)&+0.083(51)&&\\\hline
}
{Results for $\Delta q$ obtained with~(R$^{\rm opt}$) and without~(R)
the use of variance reduction techniques.
\label{tab2}}

\section{Results and Outlook}
\FIGURE
{\rotatebox{270}{\includegraphics[height=.44\textwidth,clip]{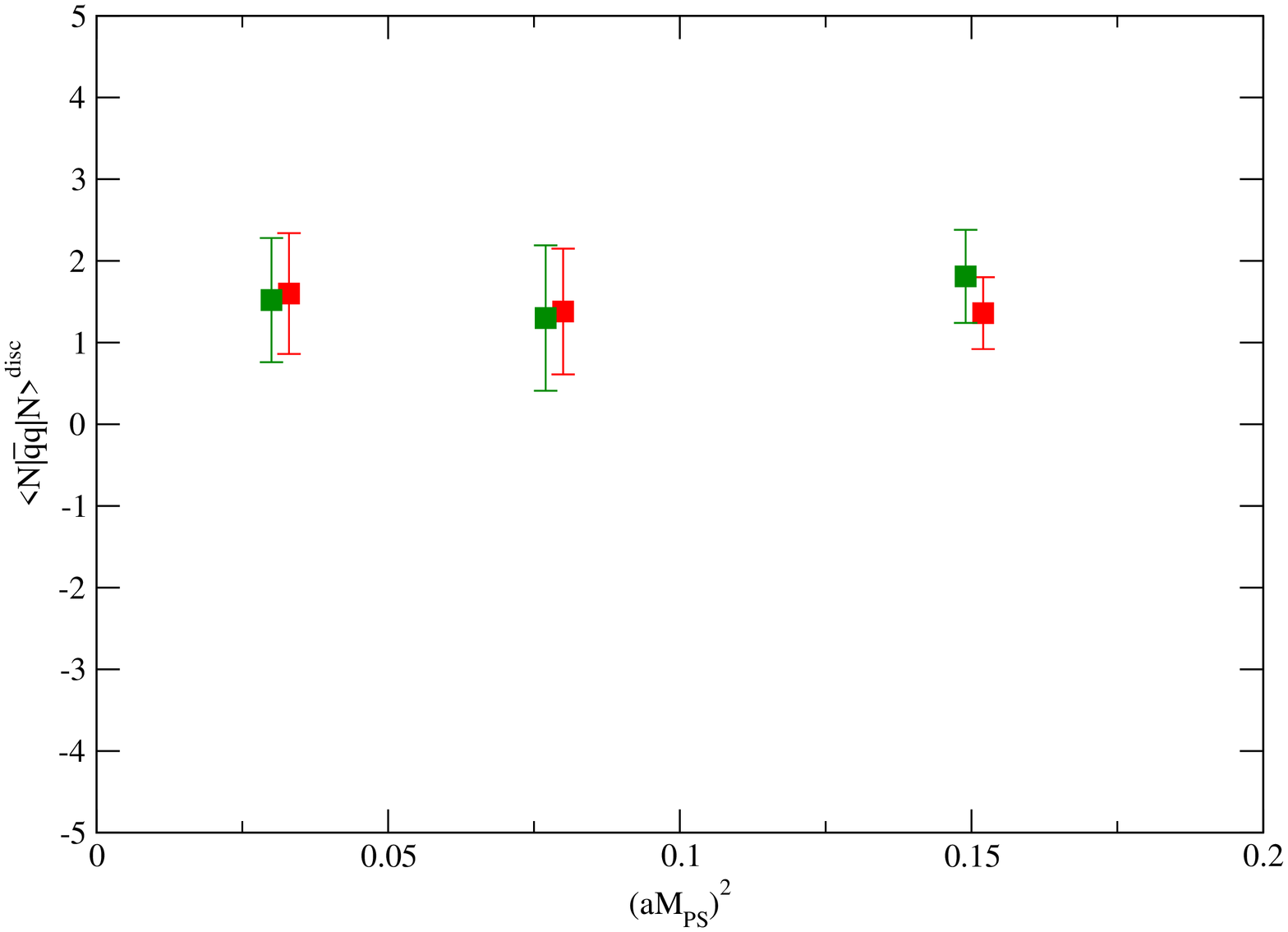}}
\hspace*{0.05\textwidth}
\rotatebox{270}{\includegraphics[height=.46\textwidth,clip]{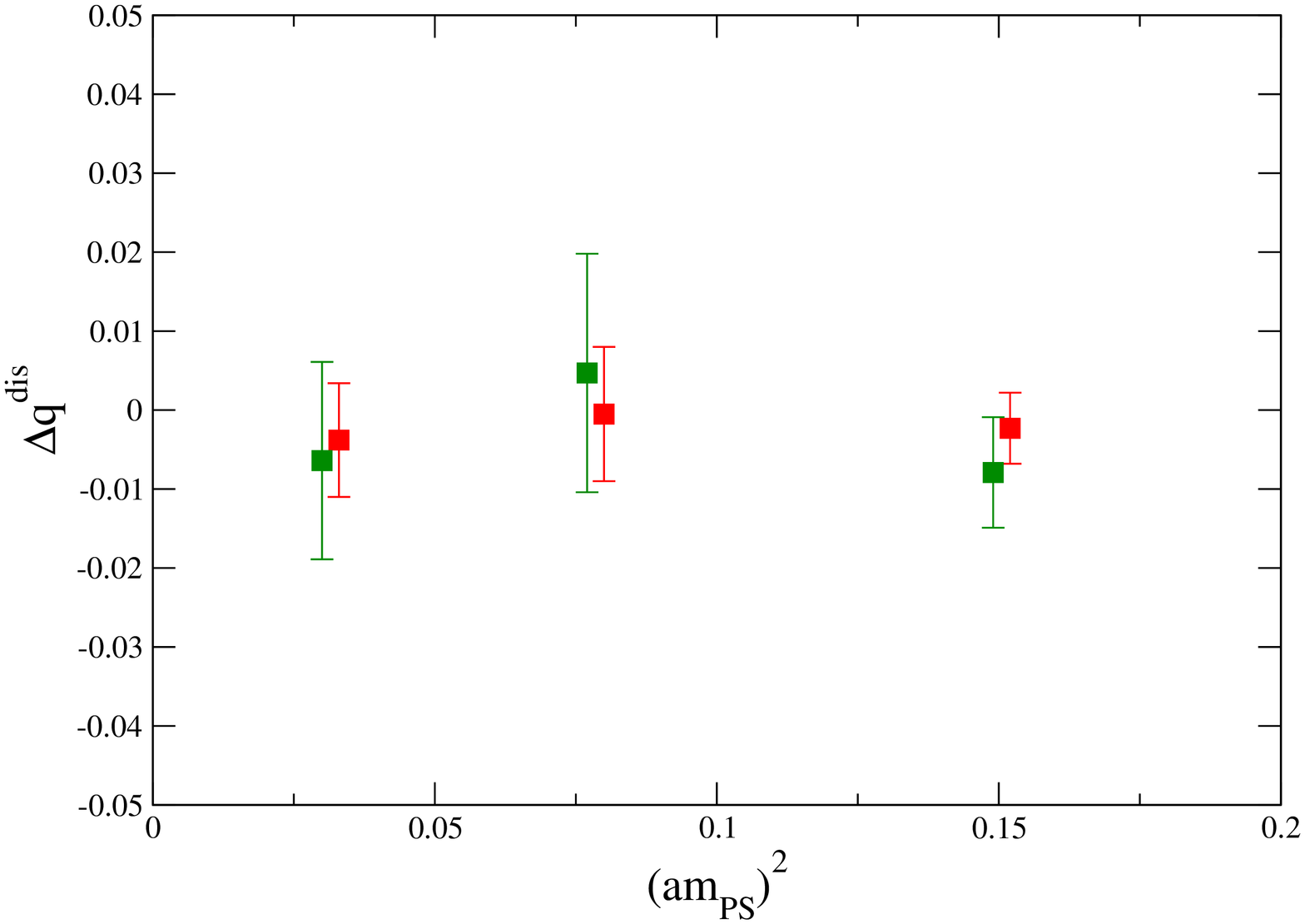}}
\caption{$\langle N|\bar{q}q|N\rangle^{\rm dis}$~(left) and $\Delta
  q^{\rm dis}$~(right) as functions of the quark mass used in the
  disconnected loop~(expressed in terms of $aM_{PS}^2$).
 The green points corresponds to a proton with
  $\kappa_{\rm 2pt}=0.1675$, while for the red points $\kappa_{\rm 2pt}=0.166$.
\label{fig3}}}
In figure~\ref{fig3} we display our results for the two matrix elements
where we obtained $\langle N|\bar{q}q|N\rangle^{\rm dis}$ at the cost
of 12 solver applications per configuration and $\Delta q^{\rm dis}$
at the cost of 100 applications, in addition to the 12 applications that
are necessary to
calculate the two point functions. In neither case do we
observe
any significant dependence on the valence quark mass, varying this from
$m_{\pi}\approx 600$~MeV down to 450~MeV, or on the 
loop quark mass, reducing $m_{\pi}\approx 600$~MeV
($\simeq$ strange quark mass) to $m_{\pi}\approx 300$~MeV.
We find 
$|\Delta s|<0.011$ at the heavier proton mass and
$|\Delta s|<0.022$ at the lighter mass value 
with 95~\% confidence level while the scalar
matrix element appears to be somewhat larger than {\em one}.
Note however that the lattice results presented here are
unrenormalized.

In the near future we will further reduce the quark masses
and the statistical errors,
in particular also of the scalar density, by refining our methods.
We will also move to non-perturbatively
improved Wilson sea quarks, allowing us to
renormalize the results and to obtain a well-defined continuum
limit.

\acknowledgments
We thank Z.\ Fodor and K.\ Szabo for providing us with the gauge
configurations.
S.~Collins  acknowledges support from the
Claussen-Simon-Foundation (Stifterverband f\"ur die Deutsche
Wissenschaft). This work was supported by the DFG
Sonderforschungsbereich/Transregio 55.

\end{document}